# Electric fluid to electric current: The problematic attempts of abstraction to concretization

**Gerald Tembrevilla, Marina Milner-Bolotin, Stephen Petrina**

Curriculum and Pedagogy, University of British Columbia, Vancouver, Canada

E-mail: gerald.tembrevilla@alumni.ubc.ca

## Abstract

In ancient times, electricity was viewed as a spirit or soul residing in an inanimate substance. A fluid image was introduced in an attempt to visualize its movement. It was also an effort to concretize its nature into the realm of physical objects as opposed to metaphysical. Eventually, a water circuit analogy has evolved and been introduced in science and engineering textbooks, old and new, to extend the depiction of a 'fluid image'. However, educational research in physics and science education has documented a number of pedagogical issues regarding the use of this analogy.

This paper outlines the historical attempts in the concretization of a 'fluid image' to promote History of Science approach in teaching physics among physics teachers in first year undergraduate and physics in-service teachers. Equally important, it discusses the pedagogical limitations of this analogy as it relates to science teaching and learning processes in secondary physics education. The discussions might serve as pedagogical considerations for physics instructors that need to connect with students entering post-secondary physics courses.

Keywords: history, electric fluid, electric current, water flow analogy, first year undergraduate physics education

## 1. Introduction

Tracing the history of science of electricity is a never-ending search of which-comes-first events. Finding one piece of history brings an awareness that precursor consequential events exist. Searching for these events may prove endless in that even the best libraries are helpless. To anyone delving into the history of science of electricity, Mottelay [1] alluded Silvanus P. Thompson's  invitation to the readers to:

let him [her or them] read Faraday, not through the eyes of Maxwell or of Tyndall, but in his own words in the immortal pages of 'Experimental Researches' with their wealth of petty detail and their apparent vagueness of speculation…

let him read Ohm's own account of the law of circuit, not some watered-down version…

let him turn over the pages of Franklin's letters to Collinson, as his observations dropped red-hot out of the crucible of his endeavours…

let him read Stephen Gray's charming experiments in the old-world diction that befitted a pensioner of the Charter-house…Not until he has thus become a bit of an antiquary will be have fully understood how the discoveries of old were made [1].

**The Amber Effect**

The history of electricity is viewed as deeply connected with magnetism as amber is often described with iron or lodestone. Amber, when exposed to fire is attracted by a magnet [2]. It is also "coupled and confused with the attraction for iron by the lodestone" [2]. 2637 BC is considered the earliest date when history noted Hoang-Ti (Yeou-hioung-che, also named Koung-fun and Hiuen-yuen), Emperor of China who utilized something which exhibits a magnetic influence [1]. It is said that Hoang-ti [1], also called Yan-ti or Hinan-yuan, built a chariot, "indicating the South, in order to distinguish the four cardinal points", directed his troops in pursuit to capture the rebellious prince Tchéyeou (Tchi-you) by the aid of the compass [4]. Between 1033-975 BC, King Solomon of Israel, the son of King David, is claimed to have used the compass and around 1000-907 BC, Homer, considered the father of Greek Poetry, related loadstone as utilized by the Greeks for navigation during the siege of Troy [1]. The "golden glow of the polished beads suggested the beaming sun, called by Homer ἠλέκτωρ which doubtless gave rise to the Greek name for amber, ἤλεκτρον" [5].

Although amber has been found in the ancient lake dwellings of Europe and in the royal tombs at Mycenae at around 2000 BC [6], Thales of Miletus, Ionia at around 600-580 BC, is credited to have been the first to witness electrical phenomenon initiated by friction in amber [1]. Thales, including Theophrastus, Pliny, and other Greek and Roman writers stated that when amber is heated, "it will attract straws, dried leaves, and other light bodies in the same way that a magnet attracts iron" [1]. At this date, 218 AD, Salmasius stressed that among the Arabs, amber was coined as *karabe* or *kahrubd* which means the power of attracting straws in Persian language [7]. There were brief discussions on the nature of amber by Roger Bacon and Peregrinus Petrus between 1254 and 1269 as compared to lodestones and magnets [1, 8]. After over three centuries, Heyl [5] claimed that the first answer for his question 'What is electricity?' or Roller & Roller's [3] 'amber effect' was given by Jerome Cardan as described by Mottelay [1], who illustrated amber as "has a fatty and glutinous humor which, being emitted, the dry object desiring to absorb it is moved toward the source, that is the amber. For every thing, as soon as it begins to absorb moisture, is moved toward the moist source, like fire to its pasture; and since the amber is strongly rubbed, it draws the more because of its heat" [6]. Benjamin stated that this description is reasonable and the first hypothesis to finally explain the phenomenon of amber as differentiated from the lodestone. This was further strengthened by the experiments of Gilbert, who showed that many substances other than amber exhibit the same property [8]. These substances are called 'electrics', having 'amber-like properties' [9]. Succeeding events brought scientific progress in the understanding of 'electrics'. The invention of the first frictional electric machine by Otto von Guericke in 1660 made him hear the sound and see the light of artificially excited electricity from a globe of sulphur. In 1705, Hauksbee constructed an electrical machine by replacing von Guericke's sulphur globe by glass as was first used by Isaac Newton [1]. According to Heyl [5], Gray's discovery of electrical conduction and electrostatic repulsion played significant roles in speculation on the nature of electricity. Moreover, in the eighteenth century, Du Fay's (1738 and later) and Benjamin Franklin's experiments stand as major developments of the fluid theory, yet in a separate trend [5].

## 2. The Fluid Theories

The French scientist DuFay communicated to the Royal Academy of Sciences at Paris his seven discoveries on electricity by introducing 'two Classes of Electricity' [10]. According to him, when a body attracts a silk thread that was rendered 'electrical', then that body is called 'vitreous'. If the same body repels the same silk thread, then that body is called 'resinous' [10]. For example, "two Silk Ribbons when rendered electrical will repel each other; Two Woolen Threads will do the like, but a Woolen Thread and a Silk Thread will mutually attract one another" [10].

The word 'circuit' was introduced by Watson communicated to the Royal Society [11]. This circuit was described in an account to discover whether or not the electrical power, when the conductors are not connected by electrics would still be felt at great distances [11]. As specifically described, "The Circuit is here formed by the coated Phial, and Hook, and the Wire between these Persons. If these Persons stand upon, or touch with any Part of their bodies any Non-electrics, which readily conducting Electricity, the Circuit is completed" [11].

Significantly, from Watson's argument based on the results of several experiments performed by "some Gentlemen of the Society" [11], the term 'fluid' was introduced. His first statement was, "what we call Electricity is the effect of very subtil and elastic Fluid, diffused throughout all our Bodies in Contact with the Globe (those Substances hitherto termed Electrics per se probably excepted), and every-where, in its natural State of the same Degree of Density" [11]. His second argument has had similarity with DuFay's report on the attraction and repulsion of bodies for 'two Classes Electricity'. He stated that, "That this Fluid manifests itself only, when Bodies capable of receiving more thereof than their natural Quantity are properly disposed for that Purpose; and that then, by certain known Operations, its Effects show themselves by attracting and repelling light Substances…" [11]. As he continued his report, "At this time I am more particular concerning the Solution of this singular Appearance, as Mr. Collinson, a worthy Member of this Society, has received a Paper concerning Electricity from an ingenious Gentleman, Mr. Franklin, a Friend of his in Pensylvania". 'Mr. Franklin's' experiments, though the material used was different ('Mr. Franklin' had used a Tube), the solution observed has no difference in relation with a Globe as Watson [11] explained, "having observed the same Fact here".

However, if the report of DuFay and Watson [11] who collaborated with each other by introducing the phenomenon of attraction and repulsion of bodies having received more or less of their natural quantity, the paper of 'Mr. Franklin' presented 'plus' and 'minus' as a new terminology:

> Hence have arisen some new Terms among us. We say, B is electrised positively; A, negatively; or, rather, *B* is electrised *plus, A, minus*…To electrise *plus* or *minus*, no more needs be known than this, that the Parts of the Tube or Sphere that are rubbed, do in the Instant of the Friction attract the electrical Fire, and therefore take it from the Thing rubbing. The same Parts immediately, as the Friction upon them ceases, are disposed to give the Fire, they have received, to any body that has less [11].

Although Franklin's one-fluid theory has some theoretical objection, its simplicity harmonizes with Newton's theories and with his increasing reputation in the scientific community, it was favored over DuFay two-fluid theory [5]. It formed as foundation in the present 'theory of electricity' but it was said to "belong equally to Dr. Watson, for he had communicated it to the Royal Society before Franklin's opinion on the subject was known" [1].

In 1775, Alessandro Volta discovered how to develop and produce electricity in metallic bodies [1] and created a complete circuit, which facilitated the flow of the electric fluid [12]. The mathematical description of a complete circuit came into full circle when Ohm [13] published his book "The Galvanic Circuit Investigated Mathematically". It was considered as remarkable since before his time, "the quantitative circumstances of electric current had been indicated in a vague way" [13].

## 3. Electric Fluid to Electric Current to Water Flow Analogies

### *From abstract to concrete*

The ancients regarded electricity as a soul or spirit resident in an otherwise lifeless substance [5]. The Renaissance in Europe brought an attempt to concretize the nature of electricity as of 'ordinary' matter to assimilate as far as possible into the tangible order of things observed by the senses. If men could not reduce electricity to a substance, they could at least describe it as though it were one. Hence, the inevitable and significant analogies – electric and magnetic 'fluids', 'electric' current, flowing like so much water in a pipe, all directly borrowed from more tangible things [9]. In support of concretizing electricity, Cardan describing the phenomenon possessed by amber, "boldly assigns a material cause" as he asserted a "link must be physical not a metaphysical one" [6]. Consequently, it must be noted that Cardan's description of amber did not categorically mention either 'fluid' or 'electric fluid'; rather it contained verb phrases of movements. For example, phrases like *'being emitted'*, *'moved towards its source'*, and *'it draws the more because of its heat'* all solicit ideas of movement, flow or transfer. Similarly, ideas of movement were evident when Gray, who discovered electrical conduction [14], found that "he could convey the electrical virtue from the tube". Gray's term of 'electrical virtue' appears to be a material similar to Cardan's description of amber as a 'fatty and glutinous humor' or entity that by the influence of friction or another forms of heat, triggers movement. In addition, Robert Boyle [15] (1738) summarized some references on theories to explain the electrical attraction of amber, also denoting series of movements. For example, according to Boyle [15], Cabeaus, described it as light bodies emitting steams, expelling surrounding air like whirlwind and Pierre Gassendi added by describing amber emitting electrical rays, getting into the pores of a straw. In 1750, in his theory of electricity, Franklin introduced '+ and – charged bodies' as he explained that a body with + charged would give electricity to the earth and a body with a – charged would take electricity from the earth [16]. Interestingly, it seems that Cardan's 'fatty and glutinous humor', Gray's term of 'electrical virtue', Boyle's summary of references, and Franklin's '+ and – charged bodies' all exhibit movements and are of the same entity communicated only in distinctly different periods.

Moreover, Volta [12] is seen to use electric fluid and electric current as just one concept as he narrated a proof of the occurrence of electric current: "What proof [is there] more evident of the flow of the electric fluid [current]— as long as the connections of the conductors forming the circuit are maintained-than that such a current is only stopped by interrupting that connection?" [17]. On the other hand, for Ampere to integrate the relationship between current and magnet [1], he introduced an image of a little man "who swims up the wire with the electric fluid…in an invisible magnetic tube with the wire at its center" [8]. Canby [8] suggested that the image of 'current' was derived from this analogy.

Although allusion to "current" was common in the 1700s, Faraday's [18 – 21] "Experimental Researches in Electricity" and sections on the "Nature of Current" gave the term scientific authority and permanence. "By *current*," Faraday [19] wrote, "I mean anything progressive, whether it be a fluid of electricity, or two fluids moving in opposite directions, or merely vibrations, or, speaking still more generally, progressive forces". He clarified: "Whether a current of electricity be considered as depending upon the motion of a fluid of electricity or the passing of mere vibrations, still the essential idea of momentum might with propriety be retained" [21]. Faraday was somewhat ambivalent about the term and what it conveyed [22]. The term current is a derivative of the Latin *curro* or *currere*, meaning "to run" with common reference to running water.

The existence of 'water and water pipe' as used in modern analogy for a current flowing in a wire was seen in Trowbridge & Burndy Library's [23] book 'What is Electricity?'. However, the usage was an indirect analogy since the actual comparison involved was between "a pipe conveying steam or compressed air in common with a copper wire conveying an electric current" [23]. The indirect analogy went as "the same is true of a water pipe; the flow of water is greatly impeded by the constriction of the pipe" [23]. In addition, the element of 'water pressure' used as a parallel term for 'potential difference' in present time analogy between water flow model and electric circuit was apparent. As the book further explained, "there is an evidence of pressure in such pipes, and also on conductors carrying electric current" [23]. Lastly, completing the inclusion of resistance for Ohm's current-voltage-resistance relationship, the book cautioned that, "there is a development of heat on both pipes and on electric conductors, but this development is much greater in the case of electricity" [23]. In a separate version, the analogy was described as "the energy to be conveyed by the current itself within the wire, in much the same way as dynamical energy is carried by water flowing in a pipe" [24]. Modern physics and electrical engineering textbooks also introduced water flow analogy in different versions [25 – 27].

## 4. Pedagogical Issues on the use of Water – Flow analogy in teaching electricity

The bibliography on constructivist-oriented research on teaching and learning science outlines that about 64% of the studies documented are carried out in the domain of physics, only 21% in biology, and 15% in chemistry. The dominance of physics from this study is connected to the complexity and abstraction of its concepts [28]. Electricity is one of the physics domains where the teaching and learning process is problematic [29]. In fact "there is little need to tell most science teachers that electricity is a difficult subject to teach effectively" [30]. One aspect of difficulty, generally across science topics, is on students' prior conceptions. Viennot [31] describes them as 'highly robust' that they "outlive teaching which contradicts them" [32]. Another concern is that "electric circuits are abstract and students are expected to develop conceptual models of the relationship between non-observables qualities (current, p. d., resistance) in terms of other non-observables such as energy and electrons" [33]. It appears that simple circuits are not simple for primary and even to tertiary levels after all. Moreover, it also points to the supposed "ontological attributes of context-dependency of students' misconceptions of a particular concept" [34], varying effectiveness and issues on the use of analogies and modeling [35] including teachers' preparedness in modeling [36] or incorporating simulations of physics phenomena using applications like Physics Education Technology (PhET) simulations [37].

Recently, the significance of model-based instruction is highly emphasized in the formulation of the Next Generation Science Standards in the United States [38]. In England, new curriculum also specifies model-based approach by 'using scientific ideas and models to explain phenomena and developing them creatively to generate and test theories' [39]. As a concrete example, an epiSTEMe project has a module in electricity in which "analogies and models would not just be used to teach circuit ideas, but would also be examined explicitly to explore how such models can be used as thinking tools in science" [40].

There is a vast amount of research [41 – 42] showing that children come to science classes with prior conceptions. These prior conceptions are identified as 'alternative frameworks' [43]. Driver and Erickson [44] reported that 'even when pupils appear to have understood an idea or principle, they revert to alternative frameworks for their intuitions when faced with slightly novel tasks'. Although "alternative frameworks" were identified and reasons for their persistence examined, research "falls short of developing a reasonable view of *how* a student's current ideas interact with new, incompatible ideas" [32].

Several studies have been documented on the understanding of concepts in electricity among young children [45 – 47]. One common problem identified from these studies and continues to be an issue [48] in teaching electricity, and "probably the most intractable, is to assist pupils to discriminate between the concepts of electric current and electrical energy" [49]. Children's prior conception insists that electric current gets 'used up' in the bulb [49] and they could not reconcile the idea "how can the same fluid be worn out and conserved at the same time?"

[50]. On the difficulty of voltage as a concept, an attempt to develop primary voltage concept for students aged 11 – 14 years is considered unresolved [50], different approaches were explored [52 – 53] and suggested 'generative paths' to be developed in elementary electricity teaching [51].

Elementary science curriculum (i.e., England and Wales, Japan, British Columbia, Canada) calls that elementary school children should be taught in constructing simple circuits, modifying current's strength and developing idea of a complete circuit [54 – 56]. In the course of the exploratory activities like lighting the bulb, children would raise questions that textbook or teacher's discourse requires elaborate support. In this conceptual level, the approach is to use analogy to "make more accessible to learners what is essentially abstract and intangible"[54]. In utilizing analogies for a given system, the reasoning centers on how parts are viewed from the whole system [57]. This analytical reasoning relates also to the base domain (the analogy) to the target domain (the concept) of [58] structure-mapping approach. However, research exploring the importance of analogies only utilized them as a form of assessment and do not investigate students' learning in actual teaching-learning process [59].

"The use of water analogies to explain and teach some of the physical relationships involved in simple DC electrical circuits is naturally appealing" [60]. In fact, the 'water-flow model' is the most popular analogy used by teachers [49] but "few reduce them to hardware" [60]. The use of the model was evident in high school and college physics class demonstration [49, 58 – 62], introduced to grade school students [63 – 64] and to lower secondary students [33]. Similar to other analogies used in electricity, it seems that students have limited opportunities to physically interact with a prototype water-flow model. Without a visual, direct, and concrete model, students across levels will have more difficulty in generating 'anchoring and bridging analogies' [35] for simple circuits. However, in a large-scale college introductory physics course, the use of simulated equipment (PhET) proved to improve college students' mastery of physics concepts better than those who assembled real circuit with light bulbs, meters, and wires [65].

Consensus models of physics are thought to work as teaching model like the kinetic particle model. The water-flow model in electricity is also considered as a consensus model [66]. However, unlike the kinetic particle model [29], it is associated with pedagogical constraints [49, 58, 66]. A study of 33 second-year school pupils in the United Kingdom revealed that children had poor understanding of water flow [63]. Beginning students in physics do not have necessary understanding about water flow dynamics for the analogy to be productive [58]. In establishing the crucial distinction between electrical current (or charge) and electrical energy

('voltage'), "the model most likely fails because students cannot discriminate the notion of pressure or 'push' of the pump from the notion of the rate of water flow" [66]. Other researchers [50, 67] contend that flow analogies are inappropriate. Another concern for water flow and electric current pair is that students view electric current as an actual substance (*Matter*) rather than movement of some substance (*Process*) [67]. This thinking is believed to create a distinct ontological status that makes the concept of electric current difficult to learn [34]. This further complicates the differentiation between electric current and electrical energy (voltage) since research shows that "when novices are forced to think about the concept of voltage, they attempt to incorporate it into their existing materialistic interpretation of current" [67].

## 5. Summary

Early scientists wanted to concretize the idea of electricity to make it understandable . From Cardan's description of amber as "fatty and glutinous humor", Gray's "electrical virtue", Franklin's "+ and – charged bodies", Faraday's "progressive forces" to Ampere's little man "who swims up the wire with the electric fluid", the water analogy that describes a current in a wire was 'materialized'.

However, the depiction of electric current as a 'material entity' in a water-circuit analogy has been found to be a contradiction of the ontological category of scientific concepts. The issues are compounded, to the fact, that this analogy is widely popular among physics and science teachers.

The use of analogy, like the water analogy as a concrete and popular analogy for electricity brings confusions and continues to create difficulty in learning. The very purpose of making electric current "visible" is also one of the very reasons why the teaching-learning processes of electric current and electricity across secondary to post-secondary physics education is visibly problematic.